\documentclass[%
 reprint,%
superscriptaddress,
 amssymb, amsmath,%
 aip,cha,%
]{revtex4-1}
\usepackage{docs}%
\usepackage{dcolumn}
\usepackage{graphicx}
\usepackage{xcolor}
\usepackage{amsfonts}
\usepackage{amsmath}
\usepackage{amssymb}
\usepackage{amsthm}
\usepackage{bm}
\usepackage{hyperref}
\hypersetup{
  colorlinks   = true, 
  urlcolor     = blue, 
  linkcolor    = blue, 
  citecolor   = blue 
}

\date{\today}

\begin{document}

\title{Localized shear generates three-dimensional transport}

\author{Lachlan~D. Smith}
 \email{lachlan.smith@northwestern.edu}
 \altaffiliation[Now at the ]{Department of Chemical and Biological Engineering, Northwestern University, Evanston, IL 60208, USA.}
  \affiliation{ 
Department of Mechanical and Aerospace Engineering, Monash University, Clayton, VIC 3800, Australia
}%
 \affiliation{CSIRO Mineral Resources, Clayton, VIC 3800, Australia}
\author{Murray Rudman}%
\affiliation{ 
Department of Mechanical and Aerospace Engineering, Monash University, Clayton, VIC 3800, Australia
}%
\author{Daniel~R. Lester}
\affiliation{School of Engineering, RMIT University, Melbourne, VIC 3000, Australia}
\author{Guy Metcalfe}
\affiliation{CSIRO Manufacturing, Highett, VIC 3190, Australia}
\affiliation{Department of Mechanical and Product Design Engineering, Swinburne University of Technology, Hawthorn, VIC 3122, Australia}
\affiliation{School of Mathematical Sciences, Monash University, Clayton, VIC 3800, Australia}

\begin{abstract}  
Understanding the mechanisms that control three-dimensional (3D) fluid transport is central to many processes including mixing, chemical reaction and biological activity. Here a novel mechanism for 3D transport is uncovered where fluid particles are kicked between streamlines near a localized shear, which occurs in many flows and materials. This results in 3D transport similar to Resonance Induced Dispersion (RID); however, this new mechanism is more rapid and mutually incompatible with RID. We explore its governing impact with both an abstract 2-action flow and a model fluid flow. We show that transitions from one-dimensional (1D) to two-dimensional (2D) and 2D to 3D transport occur based on the relative magnitudes of streamline jumps in two transverse directions. \copyright 2017 AIP Publishing.
\\

\noindent DOI: \href{http://dx.doi.org/10.1063/1.4979666}{10.1063/1.4979666}
\end{abstract}



\maketitle

\begin{quotation}
While some mechanisms for 3D particle transport (i.e. not confined to 1D curves or 2D surfaces) have been uncovered, there is still much to be understood. We provide a general description of a novel mechanism for 3D particle transport, and demonstrate it in a generic model and a model fluid flow with periodically opening and closing valves. Under this mechanism particles are `kicked' between streamlines when they experience highly localized shear, such that after many visits to the shearing region particles can visit a large extent of the domain. We expect this 3D transport mechanism to occur in many natural and engineered systems that exhibit localized shear, including valved fluid flows, granular flows and flows with yield stress fluids. 
\end{quotation}

\section{Introduction}

Understanding the mechanisms that control mixing and transport is essential to many natural and engineered flows. Significant insights into fluid mixing have been gained by application of a dynamical systems approach, termed chaotic advection \cite{Aref, Aref2014}, to a range of flows including biological flows \cite{Lopez2001}, geo- and astro-physical flows \cite{Ngan1999,Wiggins2005}, and industrial and microfluidic flows \cite{Nguyen2005}. 

Fluid transport and mixing are well understood in 2D chiefly due to the direct analogy between 2D incompressible flows and Hamiltonian systems \cite{Aref,Ottino}. In contrast, much less is known about these mechanisms in 3D flows \cite{Wiggins}, because (a)~the Hamiltonian analogy breaks down at stagnation points \cite{Bajer}, and (b)~there is an explosion of topological complexity between 2D and 3D space. In particular, little is known about the mechanisms that drive transitions from 1D or 2D transport (where particles are confined to curves or surfaces) to 3D transport, an essential ingredient for complete mixing in 3D systems, e.g.\ plankton blooms where 3D transport alters macroscopic dynamics \cite{Sandulescu2008, Tel1}. 

One of the few clearly identified 3D transport mechanisms is Resonance Induced Dispersion (RID) \cite{Cartwright+Feingold+Piro2,Cartwright1995, Cartwright1996, Vainchtein2, Vainchtein+Abudu, Meiss}, where fully 3D transport is produced by localized jumps between 1D streamlines. This mechanism is most clearly described in terms of slowly varying `action' variables $I$ and fast `angle' variables $\theta$. RID occurs in 2-action systems (with two action and one angle variable), such that the volume-preserving map $\bm{x}' = \Phi_T(\bm{x})$ corresponding to the solution of the period-$T$ advection equation $\dot{\bm{x}} = \bm{v}\left( \bm{x} ,t \right)$ is transformed into
\begin{align} \label{eq:action_angle}
I_1' &= I_1 + \epsilon g_1(I_2,\theta), \quad 
I_2' = I_2 + \epsilon g_2(I_1',\theta), \\ 
\theta' &= \theta + \Omega(I_1',I_2') + \epsilon g_3(I_1',I_2') \!\!\!\! \mod 1, \nonumber
\end{align}
where $\epsilon \ll 1$ and $g_{1,2,3}$ respectively represent perturbations of $I_1,I_2,\theta$, corresponding to e.g.\ weakly inertial flows \cite{Moharana2013, Speetjens2}. When $\Omega(I_1,I_2)$ is irrational, particle trajectories densely fill a 1D streamline and (\ref{eq:action_angle}) reduces to a 2D Hamiltonian system by averaging over $\theta$ \cite{Vainchtein+Abudu}. However, when a fluid particle enters a resonance region where $\Omega(I_1,I_2)$ is within $\epsilon$ of a rational number with small denominator \footnote{The impact of lower-order resonances (smaller denominator) on particle transport is greater as they contribute more to the Fourier expansion of eq.~(\ref{eq:action_angle}). For full details see Cartwright \emph{et al.} \cite{Cartwright+Feingold+Piro2,Cartwright1995, Cartwright1996}.}, $\theta$ slows (in the co-rotating frame) to $\mathcal{O}(\epsilon)$: averaging breaks down, and fluid particles are no longer confined to streamlines of constant $I_1,I_2$. In the resonance region fluid particles can `jump' to a new streamline. Fully 3D transport results from multiple streamline jumps as fluid particles repeatedly visit resonance regions. 

In this study we uncover a 3D transport mechanism termed `Localized Shear Induced Dispersion' (LSID), which also causes streamline jumping. Unlike RID, LSID is driven by localized shear parallel to a surface $\mathcal{S}$, either as a sharp smooth or discontinuous deformation. Such localized smooth shears can arise in general fluids - e.g.\ from the opening and closing of valves \cite{Jones1988, Smith2016discdef} - and can become discontinuous in shear-banding materials such as colloidal suspensions, plastics, polymers, and alloys \cite{Boujlel2016, Louzguine2012, Olmsted2008}. For solid matter, highly localized shears occur in granular flows with thin flowing layers, which become discontinuous in the theoretical limit of on infinitesimal flowing layer \cite{Metcalfe1996,Ottino2000, Christov2010, Juarez2010,  Sturman2012, Park2016}.

\section{Localized shear induced dispersion}

LSID occurs in 2-action systems of the form
\begin{align} \label{eq:action_angle_2}
I_1' &= I_1  + f_1(I_2,\theta), \quad 
I_2' = I_2  + f_2(I_1',\theta),  \\ 
\theta' &= \theta + \Omega(I_1',I_2') + \epsilon g_3(I_1',I_2')\!\!\!\! \mod 1, \nonumber
\end{align}
where $f_{1,2}$ are functions that represent localized shear parallel to $I_{1,2}$. These functions are $\mathcal{O}(\epsilon)$ everywhere except near a surface $\mathcal{S}$, where they are $\mathcal{O}(1)$. Fluid particles that shadow streamlines of constant $I_{1,2}$ are pushed onto a new streamline by $f_{1,2}$ with each approach to $\mathcal{S}$, leading to fully 3D transport over many encounters. Therefore, LSID and RID are mutually exclusive; RID cannot occur near $\mathcal{S}$ as the resulting streamline jump would break the resonance.

Another distinguishing feature of LSID is the short-term predictability of each streamline jump. In LSID the relatively simple nature of $f_{1,2}$ means that each streamline jump can be predicted in the short-term. In contrast, in RID streamline jumps are highly sensitive to particle locations in the resonance region, and hence less predictable in the short-term. 

While it is not surprising that the functions $f_{1,2}$ in eq.~(\ref{eq:action_angle_2}) can yield 3D transport if they have sufficient magnitude, the description eq.~(\ref{eq:action_angle_2}) explains the underlying mechanism that drives streamline jumping in many more complex systems. For instance, the streamline jumping that occurs in tumbled granular flows is driven by localized shears in a thin flowing layer \cite{Christov2010}.

\subsection{LSID in a simple model}

\begin{figure*}[tbp]
\centering
\includegraphics[width=\textwidth]{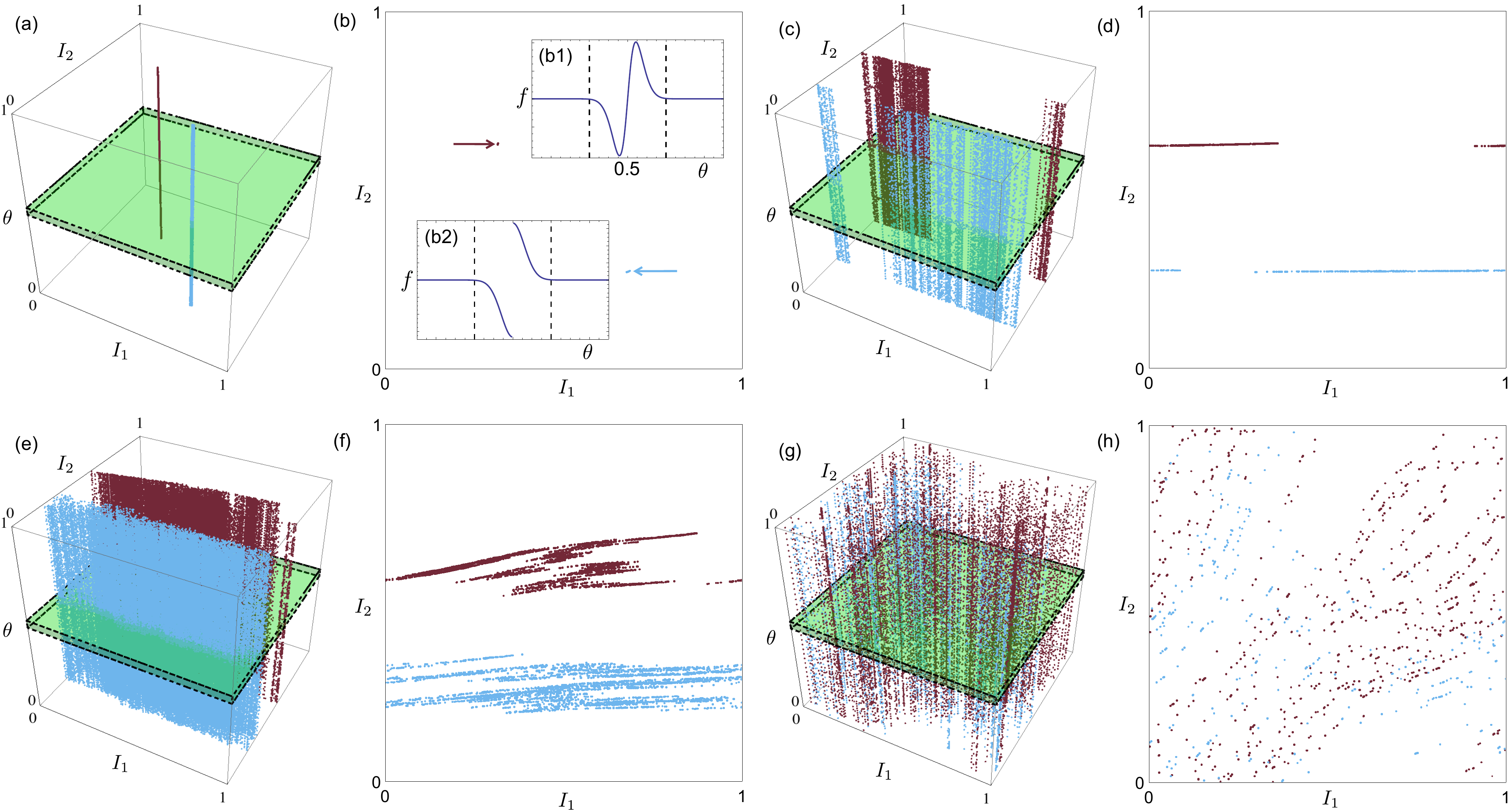}
\caption{LSID in the system described by eq.~(\ref{eq:action_angle_2}) and (\ref{eq:simple_LSID}). (a--h)~$10^5$ iterations of two tracer particles (maroon/dark gray and blue/light gray) for various choices of $a_{1,2}$. Left panels show 3D view with the region $|\theta -0.5|<\delta$ affected by the shear shown as green. Right panels show projections onto the $(I_1,I_2)$-plane. (a,b)~$a_{1,2}=5\times 10^{-4}$. (c,d)~$(a_1,a_2)=(0.05, 5\times 10 ^{-4})$. (e,f)~$(a_1,a_2)=(0.05, 5\times 10 ^{-3})$. (g,h)~$a_{1,2}=0.2$. Also shown inset in (b) is $f[(\theta-0.5)/\delta]$ from eq.~(\ref{eq:simple_LSID}) [(b1) $k=5$: smooth, (b2) $k\to \infty$: discontinuous] shown on the sub-domain $\theta\in [0.45,0.55]$. The region $|\theta -0.5|<\delta=0.02$ is shown by the dashed lines.}
\label{fig:mech_fns}
\end{figure*}

To illustrate the LSID mechanism we consider a 2-action system described by eq.~(\ref{eq:action_angle_2}) with
\begin{align} \label{eq:simple_LSID}
f_1 (I_2,\theta) &= a_1 \left(1 + \sin(2\pi I_2)/2 \right) f\left[ (\theta - 0.5)/\delta \right] \nonumber \\
f_2 (I_1',\theta) &= a_2 \left(1 + \sin(2\pi I_1')/2 \right) f\left[ (\theta - 0.5)/\delta \right]  \\
\Omega(I_1',I_2') &= (I_1' + I_2')/2, \nonumber
\end{align}
where $f_{1,2}$ are localized shears, generated by $f(x) = \tanh (k x) P(x)$ where $P(x)$ is the Gaussian distribution with mean $0$ and standard deviation $1/3$, so that $f$ is odd and close to zero outside the interval $(-1,1)$. The function $f[(\theta -0.5)/\delta]$ is shown in Fig.~\ref{fig:mech_fns}(b1,b2) for $\delta=0.02$ with $k=5$ (smooth) and $k\to \infty$ (discontinuous); it is odd about $0.5$ and rapidly decays to zero. The magnitudes of the localized shears are controlled by $a_{1,2}$, and `sharpness' is controlled by $k$. In the limit as $a_{1,2} \to 0$ the map is integrable, with particles confined to curves of constant $I_{1,2}$. When $a_{1,2}$ are both small no noticeable streamline jumps are produced, resulting in 1D transport [Fig.~\ref{fig:mech_fns}(a,b)]. Increasing only $a_1$ results in streamline jumping transverse to $I_1$ and hence 2D transport [Fig.~\ref{fig:mech_fns}(c,d)]. Increasing $a_2$ also results in streamline jumps transverse to $I_2$, and hence 3D transport [Fig.~\ref{fig:mech_fns}(e,f)]. In this case $a_1=10 a_2$ so the magnitudes of streamline jumps transverse to $I_1$ are larger than those transverse to $I_2$, resulting in slow dispersion transverse to $I_2$. Sufficiently large $a_{1,2}$ results in rapid 3D transport [Fig.~\ref{fig:mech_fns}(g,h)].


An important difference with RID is that under LSID $I_{1,2}$ `speed up' near $\mathcal{S}$ rather than $\theta$ `slowing down' near a resonant surface. This means that for LSID less time is required near $\mathcal{S}$ for streamline jumps to occur, and hence 3D transport is more rapid. This is demonstrated in the system (\ref{eq:simple_LSID}) by the rapid jumps (occurring after a single iteration) that occur, e.g.\ jumps of 3\% in $I_1$ occur immediately and almost continuously in Fig.~\ref{fig:mech_fns}(c--f). Compared to RID, the streamline jumps of LSID are faster, occur more frequently, and are more predictable in the short-term.

\subsection{LSID in a model fluid flow}

\begin{figure*}
  \begin{minipage}[c]{0.7\textwidth}
    \includegraphics[width=\textwidth]{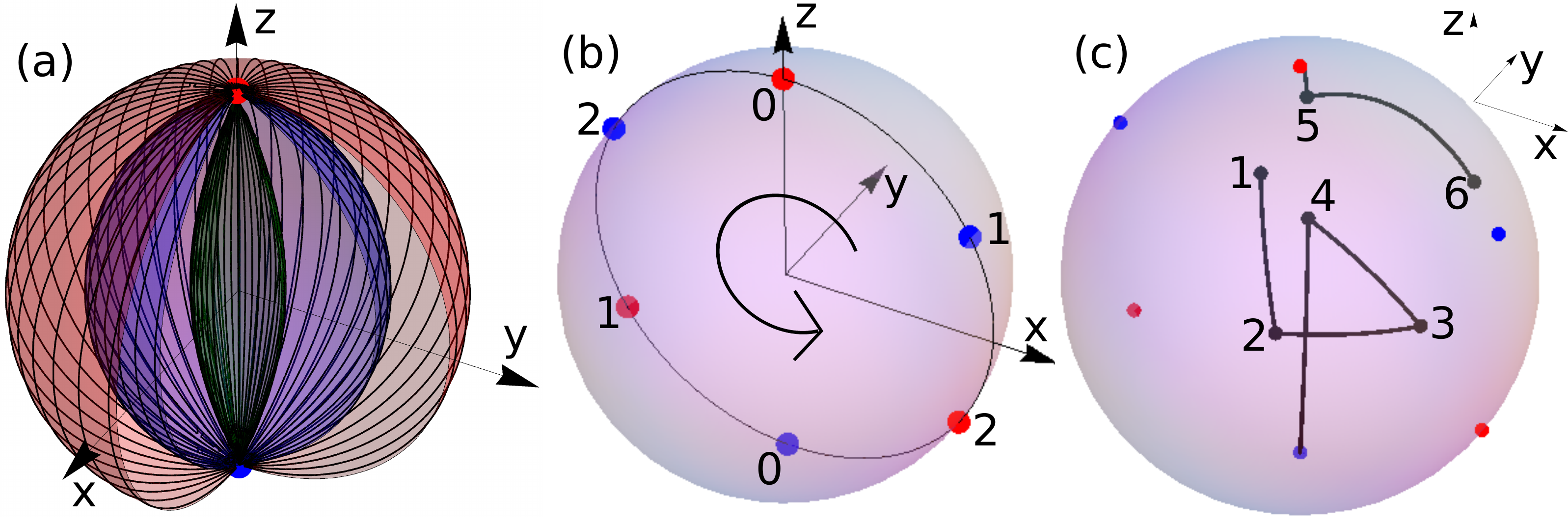}
  \end{minipage}\hfill
  \begin{minipage}[c]{0.25\textwidth}
    \caption{
      The 3DRPM flow. (a) Isosurfaces of the Stokes streamfunction $\Psi$ of the steady dipole flow. Tracer particles follow streamlines given by the solid curves with constant streamfunction and azimuthal angle. (b) The locations of the source (red) and sink (blue) pairs for the 3DRPM flow with $\Theta=2\pi/3$. (c) A typical particle trajectory with $\tau = 0.3$.
    } \label{fig:3drpm_setup}
  \end{minipage}
\end{figure*}


We now consider a model fluid flow which exhibits LSID as a result of discontinuous deformations analogous to Fig.~\ref{fig:mech_fns}(b2) rather than the previous smooth model. The 3D Reoriented Potential Mixing (3DRPM) flow \cite{Smith2016bif, mythesis, Sheldon2015}, consists of a periodically reoriented 3D dipole flow contained within the unit sphere [Fig.~\ref{fig:3drpm_setup}]. After each time period $\tau$ (where $\tau=1$ is the emptying time of the sphere) the source/sink pair is rotated about the $y$-axis by $\Theta=2\pi/3$, resulting in typical tracer particle trajectories shown in Fig.~\ref{fig:3drpm_setup}(c). Note that particles that reach the sink are immediately reinjected at the source along the same streamline, as occurs between positions 4 and 5 in Fig.~\ref{fig:3drpm_setup}(c). The steady dipole flow that is periodically reoriented is axisymmetric about the $z$-axis, it is a potential flow ($\bm{v} = \nabla \Phi$) with potential function
\begin{equation}
\Phi(\rho,z) = \frac{1}{4\pi} \left( \frac{2}{d^-} - \frac{2}{d^+} + \log \left( \frac{d^+ -z +1}{d^- +z +1} \right) \right),
\end{equation}
where $d^\pm = \sqrt{\rho^2 + (z \mp 1)^2}$ are the distances from the poles $(0,0,\pm 1)$ and $(\rho,\theta,z)$ denote cylindrical coordinates. The steady dipole flow also admits an axisymmetric Stokes streamfunction
\begin{equation}
\Psi(\rho,z) = \frac{1-\rho^2 -z^2}{4\pi} \left(\frac{1}{d^-} + \frac{1}{d^+}\right),
\end{equation}
such that $\bm{v} = (\nabla \times \Psi \hat{\bm{e}}_\theta)/\rho$, whose isosurfaces are shown in Fig.~\ref{fig:3drpm_setup}(a). The 3DRPM flow is a 3D extension of the 2D RPM flow \cite{Metcalfe2, Lester, Trefry}, which exhibits discontinuous slip deformations local to a curve of Lagrangian discontinuity due to the opening, closing and reorienting of the active dipole \cite{Smith2016discdef}. In 2D these discontinuous deformations significantly alter the classical Lagrangian dynamics that arise under smooth deformations (diffeomorphisms), leading to novel Lagrangian coherent structures \cite{Smith2016discdef}. Here we show that Lagrangian discontinuities have an even greater impact in 3D systems, providing a mechanism for 3D transport through LSID. 

\begin{figure*}
  \begin{minipage}[c]{0.7\textwidth}
    \includegraphics[width=\textwidth]{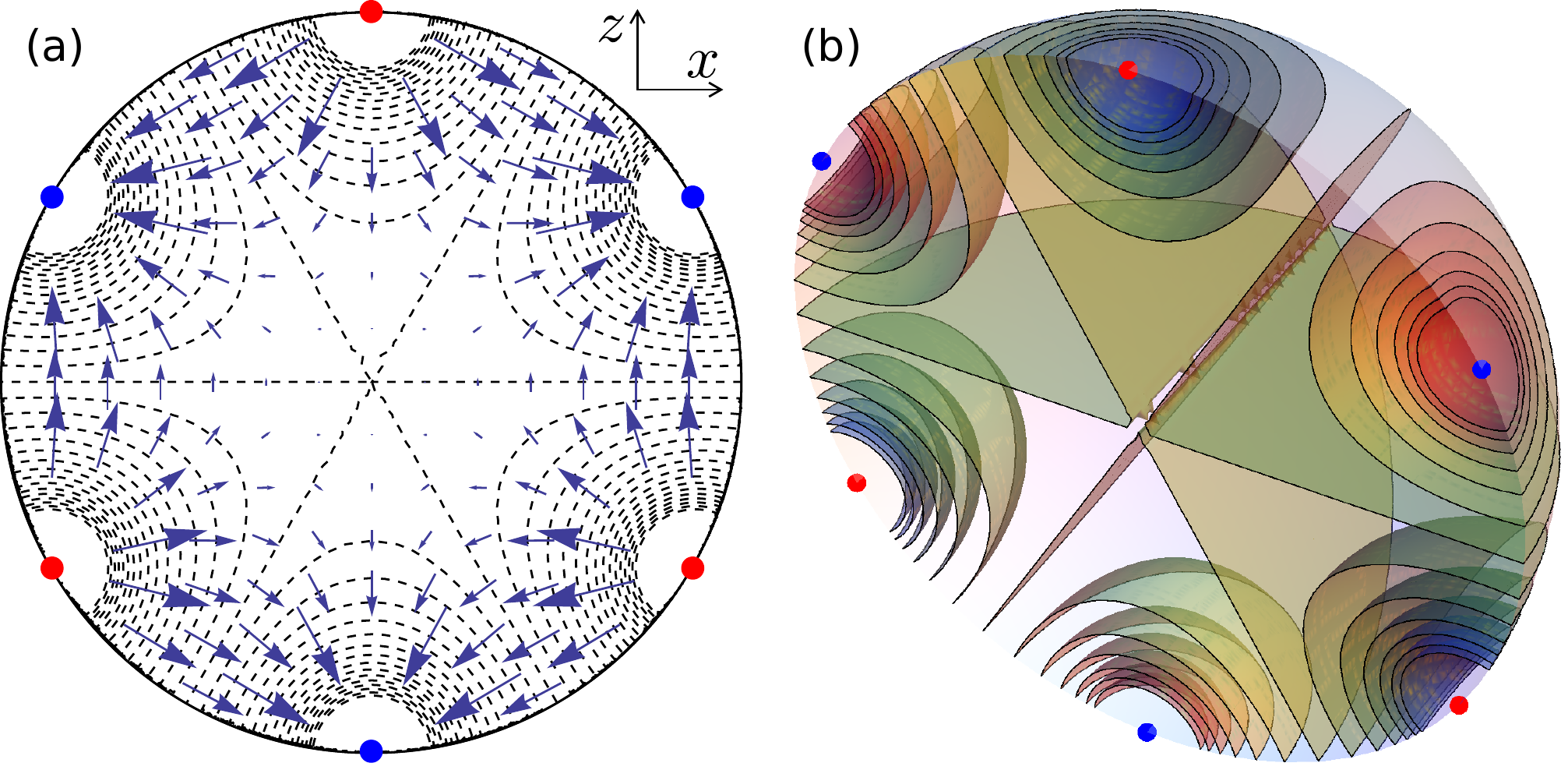}
  \end{minipage}\hfill
  \begin{minipage}[c]{0.25\textwidth}
    \caption{
      The limit as $\tau \to 0$. (a) The velocity field $\bm{v}_0$ and contours of the potential function shown in the $xz$-plane. (b) Contours of the potential function in the $y^+$ hemisphere.
    } \label{fig:3D_lim_tau_0}
  \end{minipage}
\end{figure*}


\begin{figure*}
  \begin{minipage}[c]{0.7\textwidth}
    \includegraphics[width=\textwidth]{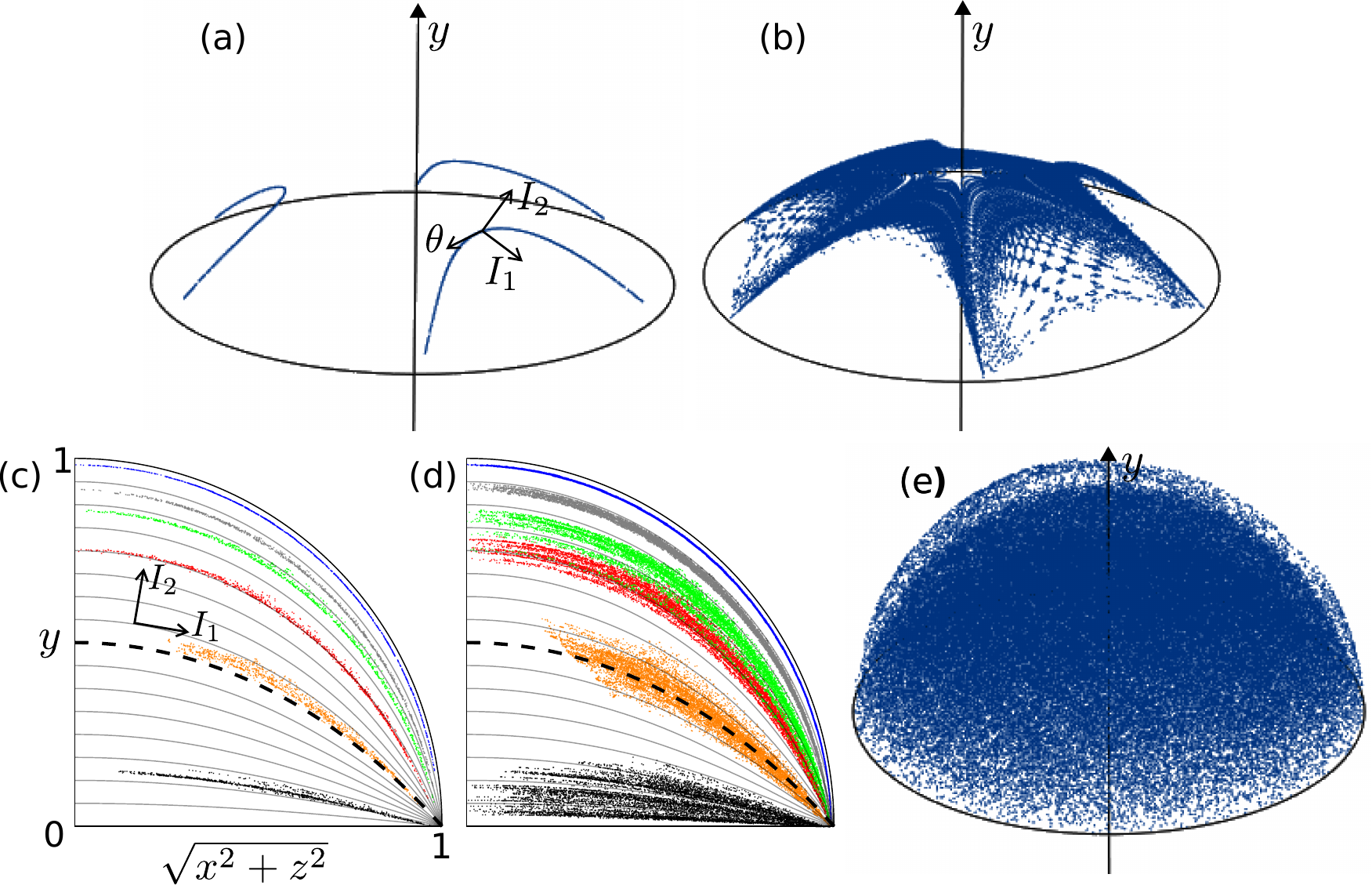}
  \end{minipage}\hfill
  \begin{minipage}[c]{0.25\textwidth}
    \caption{
      Typical Poincar\'{e} sections for the 3DRPM flow. (a)~The orbit of one tracer particle with $\tau=5\times 10^{-4}$ after $N=10^5$ flow periods. (b)~$\tau = 0.02$, $N=10^5$. (c,d)~Projections of Poincar\'{e} sections (different colors correspond to different initial positions) with $\tau=0.328$ onto $(\sqrt{x^2+z^2},y)$ coordinates. Projections of the isosurfaces of $I_2$ are shown in gray, and the isosurface $I_2=0.5$ is shown as dashed black. (c)~$N=10^3$. (d)~$N=10^4$. (e)~$\tau = 5$, $N=10^5$.
    } \label{fig:psections}
  \end{minipage}
\end{figure*}

\begin{figure*}
  \begin{minipage}[c]{0.7\textwidth}
    \includegraphics[width=\textwidth]{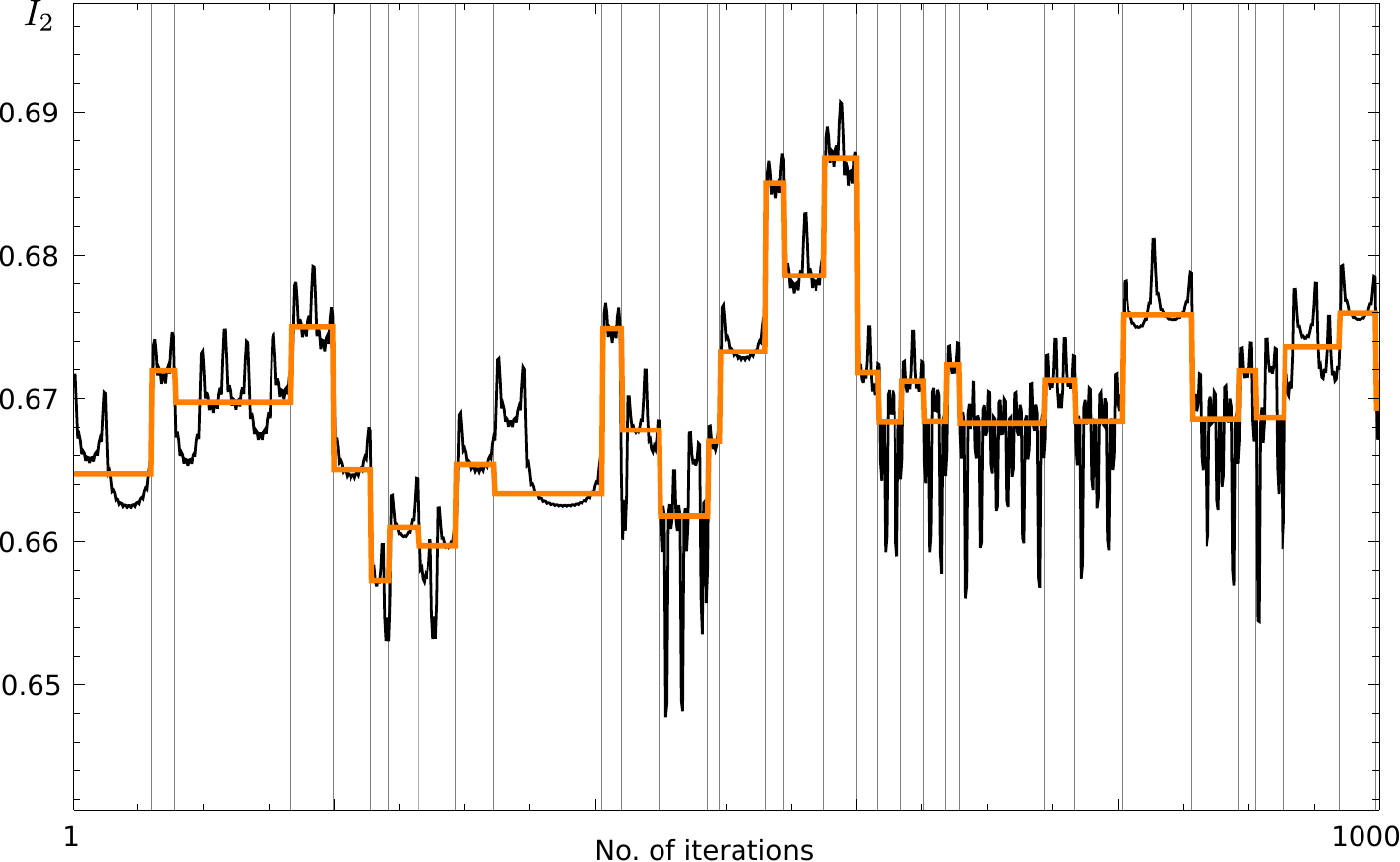}
  \end{minipage}\hfill
  \begin{minipage}[c]{0.25\textwidth}
    \caption{
      Time series of the action variable $I_2$ of a particle over 1,000 periods of the 3DRPM flow for $\tau=0.1159$ (black). The zoned median of $I_2$ (orange) with zones shown as vertical gray lines shows jumps between quasi-periodic states.
    } \label{fig:I2_time-series}
  \end{minipage}
\end{figure*}

To describe the system in terms of action-angle coordinates, we first note that in the limit as $\tau \to 0$ (infinitely fast dipole switching) the flow becomes steady and integrable, with velocity $\bm{v}_0$ given by the average of the velocity over all reoriented dipole positions [Fig.~\ref{fig:3D_lim_tau_0}]. At low values of $\tau$ ($<10^{-3}$) particles shadow the streamlines of $\bm{v}_0$, experiencing only small perturbations transverse to the streamlines [Fig.~\ref{fig:psections}(a)]. Therefore, the direction parallel to the streamlines of $\bm{v}_0$ provides a natural angle variable $\theta$ for the system, as demonstrated in Fig.~\ref{fig:psections}(a). We observe that for intermediate values, $10^{-3}<\tau<1$, particles loosely adhere to axisymmetric (about the $y$-axis) 2D surfaces, demonstrated by the 3D Poincar\'{e} section in Fig.~\ref{fig:psections}(b) and the projected Poincar\'{e} sections in Fig.~\ref{fig:psections}(c,d). These surfaces define an action variable $I_2$, and are iso-surfaces of the axisymmetric function 
\begin{equation}
I_2 = \rho_0\left( \Psi \left( y, \sqrt{x^2 + z^2} \right) \right)
\end{equation}
where $\rho_0(\psi)$ satisfies $\Psi(\rho,0) = \psi$, with equation
\begin{equation}
\rho_0(\psi) = \sqrt{ 2\pi^2 \psi^2 -2\pi \sqrt{ \pi^2 \psi^4 + 2\psi^2} + 1}.
\end{equation}
The isosurfaces of $I_2$ are shown as the gray curves in Fig.~\ref{fig:psections}(c,d), where the isosurface $I_2 = c$ intercepts the $y$-axis at $y=c$. Hence, $I_2 = 0$ corresponds to the $xz$-plane and $I_2=1$ corresponds to the outer hemisphere. The variable $I_2$ is exactly conserved by the 3DRPM flow in the $yz$ plane, however, it is not exactly conserved elsewhere, as the 3DRPM flow does not admit an invariant. While not exactly conserved, the projected Poincar\'{e} sections in Fig.~\ref{fig:psections}(c) and the relatively small deviations observed in the typical time-series shown in Fig.~\ref{fig:I2_time-series} (black curve) demonstrate that $I_2$ is approximately conserved by the 3DRPM flow for large numbers of periods at small and intermediate values of $\tau$. The other action variable, $I_1$, is defined as the direction orthogonal to $\theta$ and $I_2$, demonstrated in Fig.~\ref{fig:psections}(a,c).

\begin{figure*}[tb]
\centering
\includegraphics[width=0.9\textwidth]{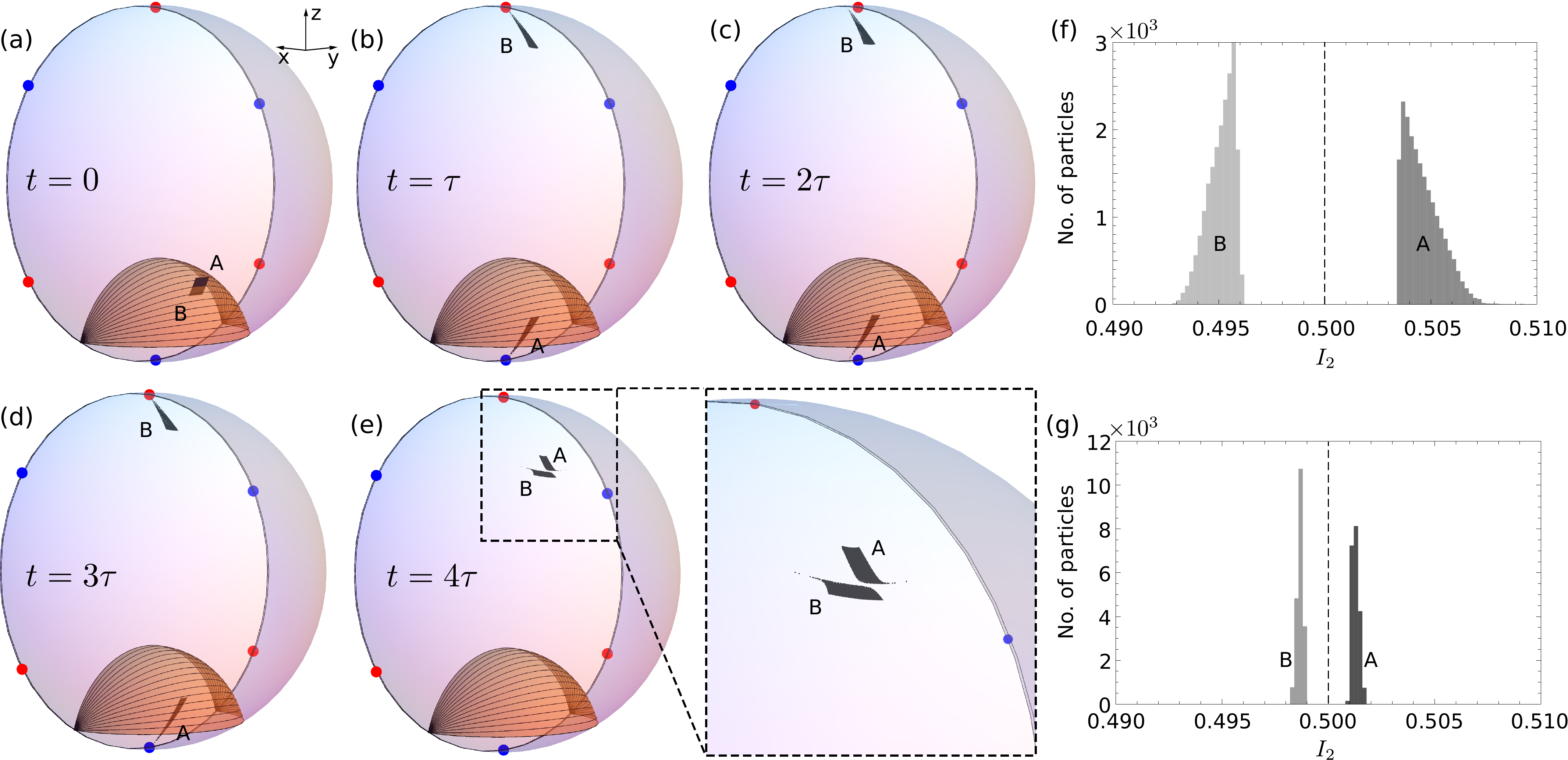}
\caption{(a--e)~The fluid cutting mechanism in the 3DRPM flow with $\tau=0.041$. (a) At $t=0$ a rectangle of black fluid lies on the isosurface $I_2=0.5$ and consists of elements A and B which straddle the orange surface consisting of locations that are advected onto the sink at $t=\tau$. (b)~At $t=\tau$ the region marked B passes through the dipole and the region marked A does not. (c--d)~The dipole is reoriented and the regions A and B move independently. (e)~At $t=4\tau$ the fluid elements A and B approximately recombine, albeit with a discontinuous slip deformation. (f)~Histograms of the action variable $I_2$ for the fluid particles A (dark gray) and B (light gray) shown in (e) at $t=4\tau$. At $t=0$ all particles were on the isosurface $I_2=0.5$ (dashed black). (g)~The same as (f) except with $\tau=0.01$.}
\label{fig:cutting_mech_3D}
\end{figure*}

Similar to the 2D RPM flow, the 3DRPM flow exhibits Lagrangian discontinuities generated by the opening and closing of the dipoles. Fluid [black rectangle AB in Fig.~\ref{fig:cutting_mech_3D}] that straddles the surface of Lagrangian discontinuity $\mathcal{S}$ [orange surface in Fig.~\ref{fig:cutting_mech_3D} given by the set of points that reach the dipole after $\tau$, i.e.\ $t_\text{sink}=\tau$] experiences a slip deformation shown in Fig.~\ref{fig:cutting_mech_3D}(a--e). This has two components that give the magnitudes of $f_{1,2}$ in eq.~(\ref{eq:action_angle_2}). In the 3DRPM flow the slip parallel to $I_1$, i.e.\ in the direction parallel to the isosurfaces of $I_2$, is generally larger and is clearly observed in Fig.~\ref{fig:cutting_mech_3D}(g). Whereas the smaller slip parallel to $I_2$ is observed via the disjoint distribution in Fig.~\ref{fig:cutting_mech_3D}(f). 

\begin{figure*}
  \begin{minipage}[c]{0.7\textwidth}
    \includegraphics[width=\textwidth]{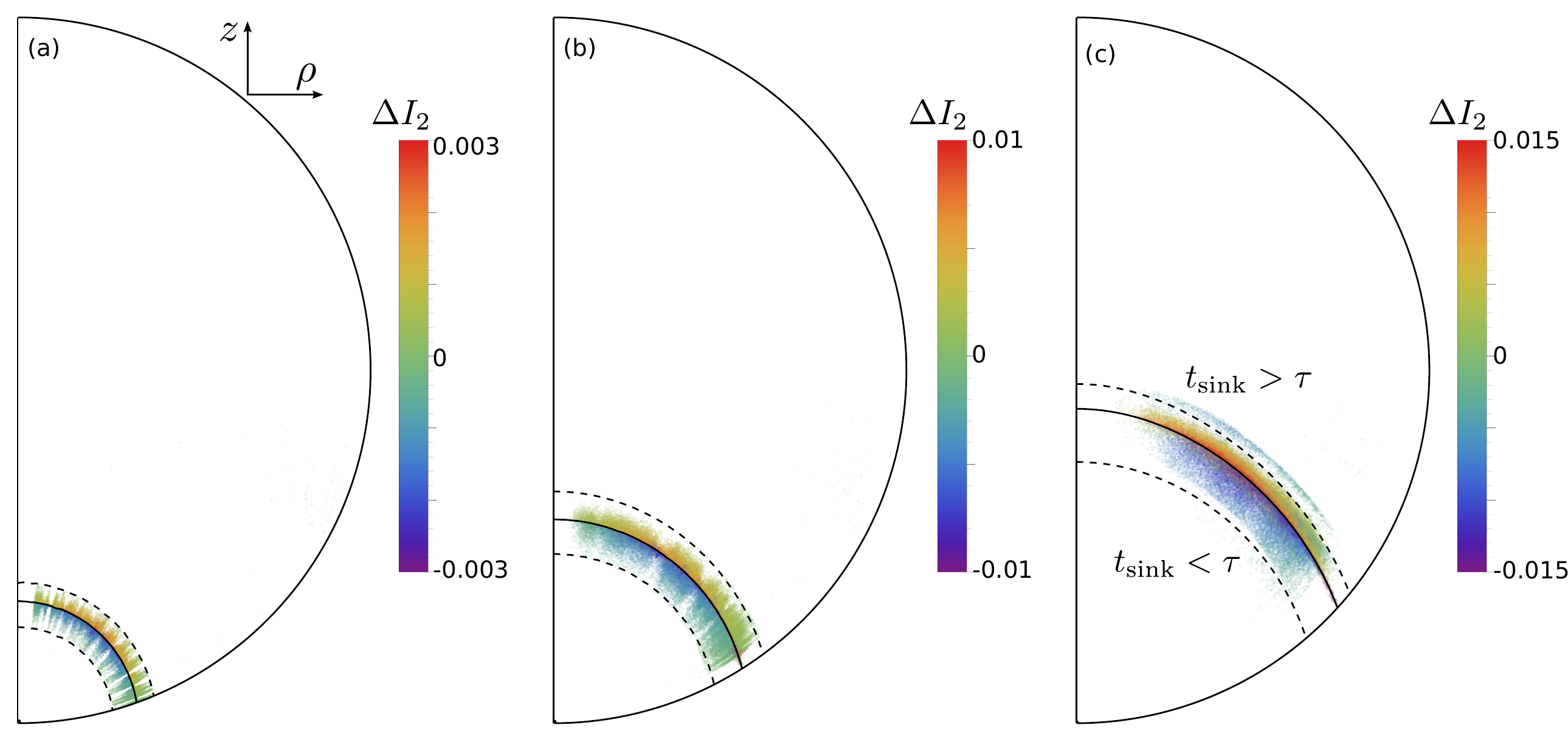}
  \end{minipage}\hfill
  \begin{minipage}[c]{0.25\textwidth}
    \caption{
     Locations of the transverse jumps $\Delta I_2$, projected onto cylindrical $(\rho,z)$-coordinates, detected by zoning time series data for (a)~$\tau = 1.024 \times 10^{-2}$, 190 initial particle locations, (b)~$\tau = 4.096 \times10^{-2}$, 58 initial particle locations, (c) $\tau=0.1159$, 12 initial particles locations. In each case each particle is advected for 200,000 flow periods. The jump locations are colored by the value of $\Delta I_2$. Also shown in each is the projected isosurface $t_\text{sink}=\tau$ (black), corresponding to the Lagrangian discontinuity (the orange surface in Fig.~\ref{fig:cutting_mech_3D}) and two projected isosurfaces of $t_\text{sink}$ that approximately contain the jump locations.
    } \label{fig:jump_locations}
  \end{minipage}
\end{figure*}


As a demonstration of the streamline jumping that can generate 3D transport, Fig.~\ref{fig:I2_time-series} shows a typical time series of $I_2$ for a particle in the 3DRPM flow with $\tau = 0.1159$. The underlying time series (black) displays jumps between quasi-periodic states, with high frequency sharp peaks occurring when the particle approaches a pole. Each quasi-periodic state indicates loose confinement to an iso-surface of $I_2$, and the jumps between them correspond to inter-surface, and hence 3D, transport. The iterations at which jumps occur are found using the data zoning method of Hawkins \cite{Hawkins1972, Hawkins1973}, dividing the dataset into segments (separated by the gray vertical lines) that are approximately piecewise constant. The magnitude of each jump, $\Delta I_2$, is found as the difference between the medians of adjacent zoned segments. Repeating this process, finding the jumps between quasi-periodic states for 12 initial particle locations, each for 200,000 flow periods, a total of 64,090 jumps were detected, and their locations in the spherical domain, shown projected onto cylindrical $(\rho,z)$-coordinates are shown in Fig.~\ref{fig:jump_locations}(c). It is evident that the inter-surface jumps occur almost exclusively near the surface of Lagrangian discontinuity (the black curve), and with greater magnitude (closer to red or purple) nearer the Lagrangian discontinuity. Considering the jump locations for two smaller values of $\tau$, Fig.~\ref{fig:jump_locations}(a,b) show that inter-surface jumps generically occur near the Lagrangian discontinuity, though with decreasing magnitude as $\tau$ decreases. Therefore, the rapid change in $I_2$ generated by the Lagrangian discontinuity is responsible for the intra-surface jumps that lead to 3D dispersion via LSID.

Like $a_{1,2}$ in eq.~(\ref{eq:simple_LSID}), $\tau$ acts as the control parameter for LSID in the 3DRPM flow, controlling the cutting mechanism demonstrated in Fig.~\ref{fig:cutting_mech_3D} and hence the magnitudes of $f_{1,2}$ in the action-action-angle description eq.~(\ref{eq:action_angle_2}). In the integrable limit as $\tau \to 0$, $I_{1,2}$ are approximately conserved. At small values of $\tau$, $f_{1,2}$ remain $\mathcal{O}(\epsilon)$ everywhere and hence particles closely shadow streamlines of $\bm{v}_0$, as demonstrated by the Poincar\'{e} section in Fig.~\ref{fig:psections}(a) which is analogous to Fig.~\ref{fig:mech_fns}(a,b). Unlike in the limit as $\tau \to 0$, particles are able to drift away from the streamlines of constant $I_{1,2}$ via small jumps when they approach the Lagrangian discontinuity. Increasing $\tau$ leads to increases in the magnitudes of both $f_1$ and $f_2$, demonstrated for $f_2$ by the increased separation between the A and B components in Fig.~\ref{fig:cutting_mech_3D}(f) compared to Fig.~\ref{fig:cutting_mech_3D}(g), and the increasing magnitude of the jumps $\Delta I_2$ in Fig.~\ref{fig:jump_locations}. As $f_1$ increases more rapidly than $f_2$, this results in approximately 2D transport [Fig.~\ref{fig:psections}(b)] analogous to Fig.~\ref{fig:mech_fns}(c,d), with rapid intrasurface transport loosely confined to isosurfaces of $I_2$. A further increase in $\tau$ results in slow dispersion away from isosurfaces of $I_2$ [Fig.~\ref{fig:psections}(c,d)] analogous to Fig.~\ref{fig:mech_fns}(e,f), and eventually 3D transport [Fig.~\ref{fig:psections}(e)] analogous to Fig.~\ref{fig:mech_fns}(g,h). Here the rapid streamline jumping produced by LSID is again illustrated, occurring after a single full flow period $3\tau$.

\begin{figure}[tb]
\centering
\includegraphics[width=0.4\textwidth]{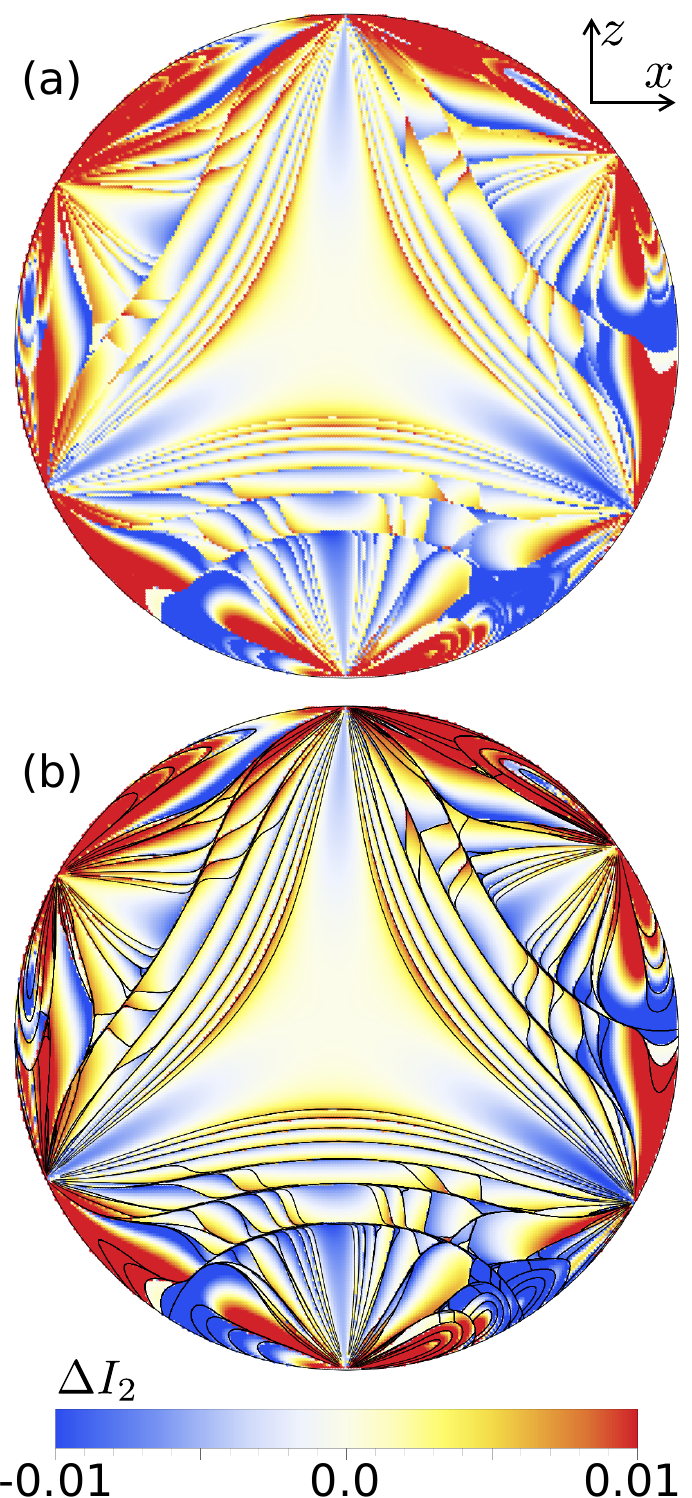}
\caption{(a) Transverse displacement of particles evenly distributed on the isosurface $I_2=0.5$ in the 3DRPM flow with $\tau=0.041$ after 20 iterations. This isosurface is projected onto the $xz$-plane to aid clarity, and each initial particle location is coloured according to the difference in $I_2$ from $0.5$ ($\Delta I_2$). (b) The same as (a) with 20 preimages (solid black curves) of the curve given by the intersection of the surface of Lagrangian discontinuity (orange surface in Fig.~\ref{fig:cutting_mech_3D}) with the isosurface $I_2=0.5$, forming part of the web of preimages of the Lagrangian discontinuity.}
\label{fig:transverse_w_web}
\end{figure}

To further highlight the dominant role that Lagrangian discontinuities play in generation of 3D transport, we track a set of fluid particles that evenly covers the isosurface $I_2=0.5$ [dashed black curve in Fig.~\ref{fig:psections}(c)]. The motion parallel to $I_2$ after 20 flow periods is shown in Fig.~\ref{fig:transverse_w_web}(a), revealing a discontinuous distribution comprised of sharp interfaces between positive and negative deviations $\Delta I_2$. These sharp interfaces correspond exactly to the `web of preimages' of the Lagrangian discontinuity (the set of reverse-time iterates of the Lagrangian discontinuity, giving locations where discontinuous deformation will occur in the future \cite {Smith2016discdef}) shown as the black curves in Fig.~\ref{fig:transverse_w_web}(b). The distribution of $\Delta I_2$ also highlights the predictable nature of the streamline jumping process over a small number of reorientations, where particles on each side of the web of preimages are driven in opposite directions by $f_{1,2}$ in (\ref{eq:action_angle_2}). In the long-term, chaotic motion can make it impossible to predict the transverse motion of particles in LSID, and the associated decorrelation allows the accumulation of jumps to be described as a diffusive-like drift, observed in Fig.~\ref{fig:psections}(c--e).

\section{Discussion}

We have shown that 3D transport can arise via LSID in flows with either smooth or discontinuous deformations in the presence of localized shears that kick particles between streamlines. These deformations can be linked theoretically by regarding discontinuous deformation as the limit of an increasingly sharp smooth deformation, as demonstrated by $f$ in (\ref{eq:simple_LSID}) which is smooth for finite $k$ but discontinuous in the limit $k \to \infty$ [Fig.~\ref{fig:mech_fns}(b2)]. For switched fluid flows with valves this connection between smooth and discontinuous deformations can be observed by considering different boundary conditions. By replacing the free-slip boundary conditions in the 3DRPM flow with no-slip conditions, fluid would not be cut at the dipole and would remain connected by a thin filament. This means the discontinuous slip deformation is replaced by a localized smooth shear. In either case LSID will be the primary source of 3D particle transport.

\section{Conclusions}

Localized shears that occur in a wide array of applications including valved flows, granular flows, and shear-banding materials can produce LSID, a mechanism for 3D particle transport. Under LSID, fluid particles are pushed to a new streamline with each pass through a localized region with high shear, leading to fully 3D transport over the ergodic region after many passes near this surface. In both a 2-action model and a valved flow transitions from 1D to 2D transport and 2D to 3D transport result from transitions in the magnitude of streamline jumps in two transverse directions as control parameters increase. 

Further investigation is required to determine how `sharp' shears need to be to produce LSID. Future studies should also focus on other 2-action systems with localized shears that are likely to experience LSID, to gain a better understanding of the possible transport and mixing phenomena it can produce. For instance, does the rate and type (diffusive, sub-diffusive etc.) of transverse particle drift depend on the nature (e.g.\ discontinuous, smooth) of the functions $f_1,f_2$ in eq.~(\ref{eq:action_angle_2})?

\begin{acknowledgments}
L. Smith was funded by a Monash Graduate Scholarship and a CSIRO Top-up Scholarship.
\end{acknowledgments}


\begin{thebibliography}{10}

\bibitem{Aref}
H.~Aref.
\newblock Stirring by chaotic advection.
\newblock {\em J. Fluid Mech.}, 143:1--21, 1984.

\bibitem{Aref2014}
H.~Aref, J.~R. Blake, M.~Budi{\v{s}}i{\'c}, J.~H. Cartwright, H.~J. Clercx,
  U.~Feudel, R.~Golestanian, E.~Gouillart, Y.~L. Guer, G.~F. van Heijst, et~al.
\newblock Frontiers of chaotic advection.
\newblock {\em arXiv preprint arXiv:1403.2953}, 2014.

\bibitem{Bajer}
K.~Bajer.
\newblock Hamiltonian {Formulation of the Equations of Streamlines in
  Three-dimensional Steady Flows}.
\newblock {\em Chaos, Solitons \& Fractals}, 4:895--911, 1994.

\bibitem{Boujlel2016}
J.~Boujlel, F.~Pigeonneau, E.~Gouillart, and P.~Jop.
\newblock Rate of chaotic mixing in localized flows.
\newblock {\em Phys. Rev. Fluids}, 1:031301, Jul 2016.

\bibitem{Cartwright+Feingold+Piro2}
J.~H.~E. Cartwright, M.~Feingold, and O.~Piro.
\newblock Passive scalars and three-dimensional {Liouvillian} maps.
\newblock {\em Physica D}, 76:22--33, 1994.

\bibitem{Cartwright1995}
J.~H.~E. Cartwright, M.~Feingold, and O.~Piro.
\newblock Global diffusion in a realistic three-dimensional time-dependent
  nonturbulent fluid flow.
\newblock {\em Phys. Rev. Lett.}, 75:3669--3672, Nov 1995.

\bibitem{Cartwright1996}
J.~H.~E. Cartwright, M.~Feingold, and O.~Piro.
\newblock Chaotic advection in three-dimensional unsteady incompressible
  laminar flow.
\newblock {\em J. Fluid Mech.}, 316:259--284, 1996.

\bibitem{Christov2010}
I.~C. Christov, J.~M. Ottino, and R.~M. Lueptow.
\newblock Streamline jumping: A mixing mechanism.
\newblock {\em Physical Review E}, 81(4):046307, 2010.

\bibitem{Hawkins1973}
D.~Hawkins and D.~Merriam.
\newblock Optimal zonation of digitized sequential data.
\newblock {\em Mathematical Geology}, 5(4):389--395, 1973.

\bibitem{Hawkins1972}
D.~M. Hawkins.
\newblock On the choice of segments in piecewise approximation.
\newblock {\em IMA Journal of Applied Mathematics}, 9(2):250--256, 1972.

\bibitem{Jones1988}
S.~W. Jones and H.~Aref.
\newblock Chaotic advection in pulsed source-sink systems.
\newblock {\em Phys. Fluids}, 31:469--485, 1988.

\bibitem{Juarez2010}
G.~Juarez, R.~M. Lueptow, J.~M. Ottino, R.~Sturman, and S.~Wiggins.
\newblock Mixing by cutting and shuffling.
\newblock {\em EPL}, 91(2):20003, 2010.

\bibitem{Lester}
D.~R. Lester, G.~Metcalfe, M.~G. Trefry, A.~Ord, B.~Hobbs, and M.~Rudman.
\newblock Lagrangian topology of a periodically reoriented potential flow:
  {Symmetry}, optimization, and mixing.
\newblock {\em Phys. Rev. E}, 80:036208, 2009.

\bibitem{Lopez2001}
C.~L{\'o}pez, Z.~Neufeld, E.~Hern{\'a}ndez-Garc{\'\i}a, and P.~H. Haynes.
\newblock Chaotic advection of reacting substances: Plankton dynamics on a
  meandering jet.
\newblock {\em Phys. Chem. Earth Pt. B}, 26(4):313--317, 2001.

\bibitem{Louzguine2012}
D.~V. Louzguine-Luzgin, L.~V. Louzguina-Luzgina, and A.~Y. Churyumov.
\newblock Mechanical properties and deformation behavior of bulk metallic
  glasses.
\newblock {\em Metals}, 3(1):1--22, 2012.

\bibitem{Meiss}
J.~D. Meiss.
\newblock The destruction of tori in volume-preserving maps.
\newblock {\em Commun. Nonlinear Sci. Numer. Simulat.}, 17:2108--2121, 2012.

\bibitem{Metcalfe2}
G.~Metcalfe, D.~R. Lester, A.~Ord, P.~Kulkarni, M.~Rudman, M.~G. Trefry,
  B.~Hobbs, K.~Regenauer-Lieb, and J.~Morris.
\newblock A partially open porous media flow with chaotic advection: towards a
  model of coupled fields.
\newblock {\em Phil. Trans. R. Soc. A}, 368:217--230, 2010.

\bibitem{Metcalfe1996}
G.~Metcalfe and M.~Shattuck.
\newblock Pattern formation during mixing and segregation of flowing granular
  materials.
\newblock {\em Physica}, A233:709--717, 1996.

\bibitem{Moharana2013}
N.~R. Moharana, M.~F.~M. Speetjens, R.~R. Trieling, and H.~J.~H. Clercx.
\newblock Three-dimensional {Lagrangian} transport phenomena in unsteady
  laminar flows driven by a rotating sphere.
\newblock {\em Physics of Fluids}, 25(9):093602, 2013.

\bibitem{Ngan1999}
K.~Ngan and T.~G. Shepherd.
\newblock A closer look at chaotic advection in the stratosphere. part i:
  Geometric structure.
\newblock {\em J. Atmos. Sci.}, 56(24):4134--4152, 1999.

\bibitem{Nguyen2005}
N.-T. Nguyen and Z.~Wu.
\newblock Micromixers -- a review.
\newblock {\em J. Micromech. Microeng.}, 15(2):R1, 2005.

\bibitem{Note1}
The impact of lower-order resonances (smaller denominator) on particle
  transport is greater as they contribute more to the Fourier expansion of
  eq.~(\ref {eq:action_angle}). For full details see Cartwright \protect \emph
  {et al.} \cite {Cartwright+Feingold+Piro2,Cartwright1995, Cartwright1996}.

\bibitem{Olmsted2008}
P.~D. Olmsted.
\newblock Perspectives on shear banding in complex fluids.
\newblock {\em Rheol. Acta}, 47(3):283--300, 2008.

\bibitem{Ottino2000}
J.~Ottino and D.~Khakhar.
\newblock Mixing and segregation of granular materials.
\newblock {\em Annu. Rev. Fluid Mech.}, 32:55--91, 2000.

\bibitem{Ottino}
J.~M. Ottino.
\newblock {\em The Kinematics of Mixing: Stretching, Chaos, and Transport}.
\newblock Cambridge University Press, 1989.

\bibitem{Park2016}
P.~P. Park, P.~B. Umbanhowar, J.~M. Ottino, and R.~M. Lueptow.
\newblock Mixing with piecewise isometries on a hemispherical shell.
\newblock {\em Chaos}, 26(7):073115, 2016.

\bibitem{Sandulescu2008}
M.~Sandulescu, C.~L\'{o}pez, E.~Hern\'{a}ndez-Garc\'{i}a, and U.~Feudel.
\newblock Biological activity in the wake of an island close to a coastal
  upwelling.
\newblock {\em Ecological Complexity}, 5(3):228 -- 237, 2008.

\bibitem{Sheldon2015}
H.~A. Sheldon, P.~M. Schaubs, P.~K. Rachakonda, M.~G. Trefry, L.~B. Reid, D.~R.
  Lester, G.~Metcalfe, T.~Poulet, and K.~Regenauer-Lieb.
\newblock Groundwater cooling of a supercomputer in {Perth, Western Australia:}
  hydrogeological simulations and thermal sustainability.
\newblock {\em Hydrogeology Journal}, 23(8) pages 1831--1849, 2015.

\bibitem{mythesis}
L.~D. Smith.
\newblock {\em Chaotic advection in a three-dimensional volume-preserving
  potential flow}.
\newblock PhD thesis, Monash University, 2016.

\bibitem{Smith2016bif}
L.~D. Smith, M.~Rudman, D.~R. Lester, and G.~Metcalfe.
\newblock Bifurcations and degenerate periodic points in a three dimensional
  chaotic fluid flow.
\newblock {\em Chaos}, 26(5):053106, 2016.

\bibitem{Smith2016discdef}
L.~D. Smith, M.~Rudman, D.~R. Lester, and G.~Metcalfe.
\newblock Mixing of discontinuously deforming media.
\newblock {\em Chaos}, 26(2):023113, 2016.

\bibitem{Speetjens2}
M.~F.~M. Speetjens, H.~J.~H. Clercx, and G.~J.~F. van Heijst.
\newblock Inertia-induced coherent structures in a time-periodic viscous mixing
  flow.
\newblock {\em Phys. Fluids}, 18:083603, 2006.

\bibitem{Sturman2012}
R.~Sturman.
\newblock The role of discontinuities in mixing.
\newblock {\em Adv. Appl. Mech.}, 45(51):51--90, 2012.

\bibitem{Tel1}
T.~T\'{e}l, A.~Moura, C.~Grebogi, and G.~K\'{a}rolyi.
\newblock Chemical and biological activity in open flows: {A} dynamical systems
  approach.
\newblock {\em Phys. Rep.}, 413:91--195, 2005.

\bibitem{Trefry}
M.~G. Trefry, D.~R. Lester, G.~Metcalfe, A.~Ord, and K.~Regenauer-Lieb.
\newblock Toward enhanced subsurface intervention methods using chaotic
  advection.
\newblock {\em J. Contam. Hydrol.}, 127:15--29, 2012.

\bibitem{Vainchtein+Abudu}
D.~L. Vainchtein and A.~Abudu.
\newblock Resonance phenomena and long-term chaotic advection in
  volume-preserving systems.
\newblock {\em Chaos}, 22:013103, 2012.

\bibitem{Vainchtein2}
D.~L. Vainchtein, J.~Widloski, and O.~Grigoriev.
\newblock Resonant mixing in perturbed action-action-angle flow.
\newblock {\em Phys. Rev. E}, 78:026302, 2008.

\bibitem{Wiggins2005}
S.~Wiggins.
\newblock The dynamical systems approach to {Lagrangian} transport in oceanic
  flows.
\newblock {\em Annu. Rev. Fluid Mech.}, 37:295--328, 2005.

\bibitem{Wiggins}
S.~Wiggins.
\newblock Coherent structures and chaotic advection in three dimensions.
\newblock {\em J. Fluid Mech.}, 654:1--4, 2010.

\end{thebibliography}
\end{document}